# Accelerating multiparametric quantitative MRI using self-supervised scan-specific implicit neural representation with model reinforcement

Ruimin Feng[a,b], Albert Jang[a,b], Xingxin He[a,b], and Fang Liu[a,b]

[a] Athinoula A. Martinos Center for Biomedical Imaging, Massachusetts General Hospital, Charlestown, Massachusetts, United States

[b] Harvard Medical School, Boston, Massachusetts, United States

# ABSTRACT

**Purpose:** To develop a self-supervised scan-specific deep learning framework for reconstructing accelerated multiparametric quantitative MRI (qMRI).

**Methods:** We propose REFINE-MORE (REference-Free Implicit NEural representation with MOdel REinforcement), combining an implicit neural representation (INR) architecture with a model reinforcement module that incorporates MR physics constraints. The INR component enables informative learning of spatiotemporal correlations to initialize multiparametric quantitative maps, which are then further refined through an unrolled optimization scheme enforcing data consistency. To improve computational efficiency, REFINE-MORE integrates a low-rank adaptation strategy that promotes rapid model convergence. We evaluated REFINE-MORE on accelerated multiparametric quantitative magnetization transfer imaging for simultaneous estimation of free water spin-lattice relaxation, tissue macromolecular proton fraction, and magnetization exchange rate, using both phantom and in vivo brain data.

**Results:** Under $4\times$ and $5\times$ accelerations on in vivo data, REFINE-MORE achieved superior reconstruction quality, demonstrating the lowest normalized root-mean-square error and highest structural similarity index compared to baseline methods and other state-of-the-art model-based and deep learning approaches. Phantom experiments further showed strong agreement with reference values, underscoring the robustness and generalizability of the proposed framework. Additionally, the model adaptation strategy improved reconstruction efficiency by approximately fivefold.

**Conclusion:** REFINE-MORE enables accurate and efficient scan-specific multiparametric qMRI reconstruction, providing a flexible solution for high-dimensional, accelerated qMRI applications.

**Keywords:** self-supervised deep learning, implicit neural representation, model reinforcement, multiparametric quantitative MRI, quantitative magnetization transfer

## 1. INTRODUCTION

Quantitative magnetic resonance imaging (qMRI) enables estimation of tissue biophysical parameters that reflect tissue composition, microstructure, and microenvironment.[1] Multiparametric qMRI allows the acquisition and reconstruction of multiple parameter maps within a single scan, substantially improving scanning workflow and enabling more comprehensive tissue characterization.[2,3] However, multiparametric qMRI typically requires the acquisition of multiple weighted images to estimate the underlying tissue parameters, resulting in prolonged scan times and increased sensitivity to motion artifacts. To accelerate multiparametric imaging, research efforts have focused on reducing data acquisition by undersampling k-space along spatial, parametric, or both dimensions, followed by advanced reconstruction methods such as parallel imaging,[4,5] compressed sensing,[6] or a combination of both.[7] Additionally, the reconstruction can be further improved in a joint spatial and parametric space by exploring spatiotemporal correlations,[8–11] direct reconstructions that incorporate the qMRI signal model,[12] imposing low-rank or subspace constraints,[13–17] and signal dictionary matching.[18] These methods leverage data redundancy and prior knowledge to reconstruct high-fidelity quantitative maps from incomplete k-space measurements.

Deep learning-based methods have shown strong potential in improving qMRI reconstruction. In particular, supervised deep learning approaches have been explored for various qMRI applications, including T1 mapping,[19] T2 mapping,[20–24] simultaneous T1 and T2 quantification,[25] and quantitative magnetization transfer mapping.[26] These methods typically rely on high-quality reference quantitative maps derived from fully sampled data to guide network training. However, acquiring fully sampled data can be challenging, especially for multiparametric qMRI applications. As an alternative, some studies have employed synthetic data generated by Bloch simulations using anatomical digital phantoms from publicly available MRI datasets.[27] More recently, self-supervised learning strategies have gained increasing attention.[28–35] These methods do not require fully sampled reference data for training; instead, they leverage MR signal models to enforce k-space data consistency during training, enabling reference-free learning. For instance, the REference-free LAtent map eXtraction (RELAX) framework[29] and its successor[33] have demonstrated self-supervised learning for variable flip angle T1 mapping. Other studies have investigated simultaneous T1 and T2 mapping using interleaved Look-Locker acquisitions with T2 preparation pulses (3D-QALAS) within a self-supervised framework.[30,31] These methods commonly use convolutional neural networks (CNNs) to map MR images to quantitative parameter maps, while relying on MR signal models to guide this cross-domain conversion through self-supervised loss minimization. Although CNNs are efficient at capturing spatial features, they are not inherently designed to exploit the spatiotemporal correlations embedded in high-dimensional qMRI data. Modeling complex MR signal behavior involving multiple tissue parameters, such as in quantitative magnetization transfer mapping, quantitative susceptibility mapping, or diffusion modeling, remains challenging for CNN-based methods, which may struggle to fully capture the intricate spatiotemporal relationships among multiple image contrasts.

Recently, implicit neural representation (INR) has emerged as a powerful technique for modeling complex signal behavior.[36–40] Unlike CNNs, which operate directly on image data, INR models the target image contrast as a function of spatial coordinates, parameterized through an encoding module such as hash encoding[41] and a multilayer perceptron (MLP) network. This coordinate-based modeling approach offers several advantages over traditional CNN-based image representations. First, multidimensional image features can be characterized as coordinate-based functions, resulting in a smooth and continuous representation in high-dimensional feature space that supports more robust learning. Second, the coordinate-based function can be significantly compressed, achieving a highly compact representation that improves learning efficiency. Third, prior knowledge, such as spatial smoothness and structural coherence, can be seamlessly integrated into the coordinate-based function, enabling flexible and adaptive learning. Previous studies have demonstrated the advantages of INR for addressing challenging image reconstruction tasks in computed tomography,[42–44] structural MRI,[45,46] and dynamic MRI.[47–52] In the context of qMRI, these advantages also make INR a promising tool for modeling spatiotemporal MR signals, supporting robust, efficient, and flexible representations for accelerated multiparametric mapping, as demonstrated, for example, by INR-based simultaneous T1, T2, and T2* mapping (SUMMIT).[34]

In this study, we propose REFINE-MORE (REference-Free Implicit NEural representation with MOdel REinforcement), which integrates INR with MR physics-based model reinforcement to advance multiparametric qMRI reconstruction. Specifically, REFINE-MORE models the underlying weighted images and multiparametric parameter maps as coordinate-based functions, parameterized by hash encodings[41] and MLPs, providing a compact and memory-efficient representation of the entire four-dimensional (3D + parametric) data. A model reinforcement module further refines these parameter estimates by enforcing data consistency with the measured k-space data, thereby improving reconstruction accuracy and robustness. We validated the proposed method using an accelerated quantitative magnetization transfer imaging sequence[53] to simultaneously estimate tissue parameters, including free water spin-lattice relaxation ($T_1^F$), tissue macromolecular proton fraction ($f$), and magnetization exchange rate ($k_F$). Experiments included both phantom and in vivo brain data. The main contributions of this work are as follows:

1. We propose a novel self-supervised, scan-specific learning framework that combines implicit neural representation with explicit MR physics-based model reinforcement to enable robust, efficient, and flexible multiparametric qMRI reconstruction.
2. We validate the proposed method on a highly undersampled (4× and 5× accelerated) 3D quantitative magnetization transfer mapping sequence for characterizing tissue macromolecular components and structures, demonstrating its feasibility in capturing complex MR signal evolution.
3. We enhance reconstruction computational efficiency by integrating a model weight transfer mechanism and a low-rank-based fast network adaptation strategy within the framework, enabling faster convergence and significantly reduced reconstruction time.

## 2. THEORY

**Figure 1** illustrates an overview of our proposed REFINE-MORE framework, which adopts a two-stage optimization strategy for multiparametric qMRI reconstruction. In the first stage, we leveraged the concept of INR to perform the de-aliasing task from undersampled measurements and provide an informative initialization for the series of quantitative parameters. This is followed by a second stage based on an unrolled proximal gradient descent algorithm, which reinforces the underlying MR physics constraints to enhance the fidelity and accuracy of the reconstructed quantitative maps.

### 2.1 Physical model in fast quantitative MRI

In qMRI reconstruction, the undersampled k-space signal $d$, can be formulated as:

$$d = \mathbf{UFCSQ} \tag{1}$$

where $\mathbf{Q} = \{q_i\}_{i=1}^{N}$ denotes a set of quantitative tissue parameters, $q_i$ is each parameter, and $N$ represents the total number of parameters to be estimated. $\mathbf{S}$ is the MR sequence-specific signal model describing the MR signal evolution. $\mathbf{C}$ represents coil sensitivity maps for multi-channel coils, $\mathbf{F}$ is the fast Fourier transform, and $\mathbf{U}$ denotes the undersampling mask for fast acquisition.

The quantitative parameters Q can be obtained from the undersampled k-space $d$ by solving the following constrained optimization problem:

$$\underset{Q}{\mathrm{argmin}}\, g(Q) + \lambda \mathcal{R}(Q) \tag{2}$$

where $g(Q) = \frac{1}{2}\|d - \mathbf{UFCS}Q\|_2^2$ enforces data consistency with the measured k-space. $\mathcal{R}(Q)$ is the regularization term that imposes prior knowledge on the estimated quantitative maps. $\lambda$ is a weighting factor that balances the contribution of the two terms. Eq. (2) can be solved by using an iterative optimization approach such as the proximal gradient descent algorithm.[54] Like the conventional curve fitting approaches, the initial estimate $Q^{(0)} = \{q_i^{(0)}\}_{i=1}^{N}$ plays a critical role in stabilizing the optimization trajectory during the gradient descent process. With a reasonably good initial value close to the optimal solution, the algorithm reduces the likelihood of being trapped in local minimums and can improve the accuracy of the parameter estimation.

## 2.2 Implicit neural representation for learning tissue parameter initialization

In the INR-based initialization module (**Figure 1(A)**) at the first stage, we employed a representation strategy to decouple the removal of image artifacts induced by k-space undersampling and the exploitation of temporal correlation characterized by the MR signal model.

More specifically, the image artifact removal is conducted by modeling the spatiotemporally varying weighted image series $W^{(0)}$ as a continuous function of both spatial coordinates $(x, y, z)$ and a 1D imaging parameter $t$ such as flip angles, echo times, inversion recovery times, etc., depending on the specific qMRI imaging protocol:

$$W^{(0)}(x, y, z, t) = MLP_1(concat[\mathcal{H}_1(x, y, z|h_1), \mathcal{H}_2(t|h_2)]|\varphi_1) \tag{3}$$

where $h_1$, $h_2$, and $\varphi_1$ are learnable parameters in the hash encodings $\mathcal{H}_1$ and $\mathcal{H}_2$, and $MLP_1$, respectively. $concat$ is a concatenation operation. A detailed explanation of hash encoding can be found in the **Supporting Background Information**. The learnable parameters in hash encodings and the MLP can be optimized by minimizing the following loss function:

$$\mathcal{L}_1 = \|d - \mathbf{UFC}W^{(0)}(h_1, h_2, \varphi_1)\|_2^2 + \lambda \|TV(W^{(0)}(h_1, h_2, \varphi_1))\|_1 \tag{4}$$

where $W^{(0)}(h_1, h_2, \varphi_1)$ represents the full 4D weighted image volumes, which are outputs of the respective functions evaluated at all sampled locations. The operator $TV(\cdot)$ denotes total variation regularization, promoting spatial smoothness in the reconstructed images. The loss $\mathcal{L}_1$ drives the INR to recover alias-free weighted images from undersampled k-space data.

Furthermore, the exploitation of temporal correlation governed by the MR signal model **S** is conducted by directly modeling the quantitative parameters $Q^{(0)}$ as a function of 3D spatial coordinates $(x, y, z)$, expressed by:

$$Q^{(0)}(x, y, z) = MLP_2(\mathcal{H}_3(x, y, z|h_3)|\varphi_2) \tag{5}$$

where $h_3$ and $\varphi_2$ also denote learnable parameters in the hash encoding $\mathcal{H}_3$ and $MLP_2$, respectively. The learnable parameters can be optimized by minimizing the following loss function:

$$\mathcal{L}_2 = \left\| W^{(0)} - SQ^{(0)}(h_3, \varphi_2) \right\|_2^2 \tag{6}$$

where $Q^{(0)}(h_3, \varphi_2)$ represent the full 3D parameter volumes. The loss $\mathcal{L}_2$ drives the INR to predict quantitative parameter maps from the estimated weighted images in loss $\mathcal{L}_1$.

This decoupled strategy simplifies reconstruction by addressing the artifact removal and parameter estimation through two complementary objectives, which facilitates stable optimization and improves interpretability.

## 2.3 Unrolling of a learnable proximal gradient descent algorithm

In the second stage for MR physics-based model reinforcement (**Figure 1(B)**), we began with the output of the INR module as the initial parameter estimate $Q^{(0)}$ for the proximal gradient descent algorithm[54] to iteratively solve Eq. (2). In the $k^{th}$ iteration, each quantitative parameter $q_i^{(k)}$ is given by:

$$\hat{q}_i^{(k)} = q_i^{(k-1)} - \alpha_i^{(k)} \nabla g\left(q_i^{(k-1)}\right) \tag{7a}$$

$$q_i^{(k)} = \text{prox}_{\mathcal{R},\alpha_i^{(k)}}\left(\hat{q}_i^{(k)}\right) \tag{7b}$$

where $\nabla$ denotes the gradient operator with respect to $q_i^{(k-1)}$. $\hat{q}_i^{(k)}$ represents the intermediate outputs after one gradient descent step. $\alpha_i^{(k)}$ is the step size in the $k^{th}$ iteration. $\text{prox}_{\mathcal{R},\alpha_i^{(k)}}$ is the proximity operator, which is defined as:

$$\text{prox}_{\mathcal{R},\alpha_i^{(k)}}(v) = \underset{x}{\text{argmin}}\left(\mathcal{R}(x) + \frac{1}{2\alpha_i^{(k)}}\|x - v\|_2^2\right). \tag{8}$$

From Eqs. (7a) and (7b), it is evident that the proximal gradient descent algorithm decouples the data consistency term from the incorporation of prior knowledge. Similar to many recent deep learning studies,[33,55–57] we unrolled the iterative optimization process into $K$ phases and replaced the proximal operator $\text{prox}_{\mathcal{R},\alpha_i^{(k)}}$ with efficient CNNs with learnable network parameters to approximate an implicit regularization term in a self-supervised manner. More specifically, in each phase, a gradient descent update is first performed for every quantitative map $q_i$ according to Eq. (7a), to enforce data fidelity and consistency with the MR physical model. This is followed by a learned regularization step, where a CNN (e.g., a shallow U-Net[58] in our implementation) is used as an implicit proximal operator to capture latent structural features and further suppress noise or artifacts in Eq. (7b). The unrolled network is optimized by minimizing the following self-supervised loss function applied to the output of all phases:

$$\mathcal{L}_3 = \sum_{k=1}^{K} \left\| d - \mathbf{UFCS}\{CNN(\hat{q}_i^k|\varphi_{cnn})\}_{i=1}^{N} \right\|_2^2 \qquad (9)$$

where $\varphi_{cnn}$ denotes the learnable parameters in the CNN. The final quantitative maps are obtained from the last phase, i.e., $Q^{(K)} = \{q_i^{(K)}\}_{i=1}^{N}$.

## 2.4 Rapid model adaptation

It should be noted that, unlike conventional deep learning reconstruction methods where training and testing are two separate steps, in REFINE-MORE with scan-specific self-supervised learning, the reconstruction is readily concluded once the training has converged for one subject. However, with the learning involving an iterative optimization process, computation efficiency might be hindered due to the gradient backpropagation within a chain of CNNs, hash encodings, and MLPs. To ensure high computation efficiency, we developed a rapid model adaptation strategy consisting of two components.

Firstly, in the INR-based initialization module, we hypothesize that the MR signal model provides strong prior information for the temporal contrast variations; therefore, the hash encoding $\mathcal{H}_1$ dependent on the imaging parameter $t$ remains less variable across different subjects under the same MR sequence. Once the first subject is reconstructed, we froze $\mathcal{H}_1$ and only optimized the hash encodings $\mathcal{H}_2$ and $\mathcal{H}_3$ that encode spatial features from 3D spatial coordinates $(x, y, z)$ in each subject. $MLP_1$ and $MLP_2$ can also be fixed, as we found them also less variable across different subjects.

Secondly, in the physics reinforcement module, we incorporated a low-rank adaptation (LoRA)[59] to improve CNN learning efficiency. Mathematically, assuming the weight matrix in the CNN is given by $\varphi_{cnn} \in \mathbb{R}^{m \times l}$, where $m$ is the number of output channels and $l$ is the number of parameters per convolution kernel, determined by the input channels and kernel size, a low-rank decomposition can be conducted to compress the full size matrix into two smaller matrices $A \in \mathbb{R}^{m \times r}$ and $B \in \mathbb{R}^{r \times l}$, where $r$ is the rank of the decomposition, so that the update of the weight matrix can be expressed as:

$$\Delta \varphi_{cnn} = \beta \cdot AB. \tag{10}$$

where $\beta$ is a scaling factor to control the magnitude of the adaptation. It should be noted that because $r \ll \min(m, l)$, the update of $\Delta \varphi_{cnn}$ can be extremely lightweight in comparison to updating a full-size $\varphi_{cnn}$ in conventional CNN learning. In our case, once the first subject is reconstructed, for the subsequent subjects, the CNN update is conducted only on $A$ and $B$. This low-rank parameterization allows the model to efficiently capture subject-specific variations while leveraging shared CNN features across volumes, enabling fast convergence. A detailed illustration of our proposed model adaptation strategy can be found in **Supporting Information Figure S1**.

## 3. METHODS

### 3.1 Multiparametric quantitative magnetization transfer imaging

We investigated the feasibility of REFINE-MORE to reconstruct accelerated multiparametric quantitative magnetization transfer imaging using a newly developed BTS (Bloch-Siegert and magnetization Transfer Simultaneously) sequence.[53] BTS models tissue with a binary spin-bath magnetization transfer system that is composed of a "free" liquid water proton pool and a "restricted" macromolecule proton pool denoted by F and R, respectively. In data acquisition, an off-resonance saturation pulse is applied after excitation, prior to acquisition in an SPGR scheme, inducing Bloch-Siegert shift (correcting for B1+ field inhomogeneity) and magnetization transfer simultaneously. The BTS signal model can be derived analytically as:

$$S_{\text{BTS}} = \rho M_0 (1-f) \sin \alpha \, \frac{1 - (E_1^F A + E_1^R B)}{1 - (E_1^F A \cos \alpha + E_1^R E_{W0} E_W C)} e^{i\varphi_{\text{BTS}}} \tag{11}$$

where

$$A = 1 - f + fE_k - E_k E_1^R E_W E_{W0}$$
$$B = f - fE_k + E_k E_W E_{W0}$$
$$C = f + E_k - fE_k$$
$$E_1^{F[R]} = e^{-R_1^{F[R]} TR}$$

$$E_{W0} = e^{-\langle W(0)\rangle \tau_{exc}}$$

$$E_W = e^{-\langle W(\Delta_{off})\rangle \tau_{BTS}}$$

$$E_k = e^{-\frac{1}{1-f}k_R TR} = e^{-\frac{1}{f}k_F TR} = e^{-kTR}$$

$$\varphi_{BTS} = B_{1,max}^2 \int_0^{\tau_{BTS}} \frac{\left(\gamma B_{1,normalized}(t)\right)^2}{2(2\pi\Delta_{off})} dt = B_{1,max}^2 K_{BS} \tag{12}$$

$\alpha$ is the excitation flip angle resulting from the excitation pulse of width $\tau_{exc}$, $R_1^{F[R]} = 1/T_1^{F[R]}$ are the corresponding spin-lattice relaxation rates of the free and restricted macromolecule proton pools, respectively, and $f$ is the macromolecular proton fraction with respect to the total magnetization $M_0 (= M_0^F + M_0^R)$ where $M_0^F$ and $M_0^R$ are the corresponding equilibrium magnetizations of the free and restricted proton pool. $\rho$ is a scaling term, $B_{1,max}$ and $B_{1,normalized}(t)$ are the peak amplitude and amplitude modulation function normalized to 1, respectively, of the saturation pulse of width $\tau_{BTS}$ applied at the offset frequency $\Delta_{off}$. $\langle W(0) \rangle$ is the average saturation rate of the macromolecule proton pool due to irradiation applied at a frequency offset $\Delta_{off}$, given by

$$\langle W(\Delta)\rangle = \pi \frac{1}{\tau_{RF}} \int_0^{\tau_{RF}} \omega_1^2(t) dt\, G_{SL}(\Delta) \tag{13}$$

$$G_{SL}(\Delta) = \int_0^1 \sqrt{\frac{2}{\pi} \frac{T_2^R}{|3u^2-1|}} e^{-2\left[\frac{2\pi\Delta T_2^R}{3u^2-1}\right]^2} du \tag{14}$$

The complexity of the magnetization transfer signal model is further manifested by the exchange of longitudinal magnetization between the two pools characterized by pseudo-first-order rate constants $k_F$ and $k_R$, where $k_F$ is the spin transfer rate from F to R, and $k_R$ is vice versa. In equilibrium, $k_F = k M_0^R$ and $k_R = k M_0^F$, where $k$ is the fundamental exchange rate constant between pools F and R. The detailed sequence protocol and signal model derivation can be found in the original BTS paper.[53]

We are primarily targeting three quantitative magnetization transfer parameters including free pool spin-lattice relaxation ($T_1^F$), macromolecular fraction ($f$), and magnetization exchange rate ($k_F$) quantification, as those parameters have been shown capable of characterizing tissue integrity in different diseases, such as demyelination in the brain,[60] and extracellular matrix degradation in the connective tissue.[61] In the original BTS study, fully sampled high-resolution quantitative magnetization transfer parameter maps have been obtained for covering the whole brain in the healthy subject with an approximate 40 min scan time, which requires imaging acceleration for clinical translation.

### 3.2 Image datasets

We collected both fully sampled in vivo and phantom data using the BTS sequence. The in vivo human brain data were acquired from five healthy volunteers in accordance with a protocol approved by the local institutional review board. Imaging was performed on a Siemens 3T Prisma scanner equipped with a 20-channel head coil. The BTS scans used the following 3D acquisition parameters: matrix size=176×176×48 with a resolution of 1.3×1.3×3.5 mm³ along the sagittal orientation, TE/TR=12/40 ms, and flip angles of 5°, 10°, 20°, and 40° with a fixed 500 $\mu s$ pulse length. For the off-resonance saturation, a Fermi pulse of 8 ms duration was applied, with a peak $B_{1,\text{max}}$ of 7.3 $\mu T$ and a frequency offset of 4000 Hz.

The phantom data were acquired on the same scanner used in the in vivo experiment. The phantom consisted of four 20 mL vials immersed in a 200 $\mu M$ $MnCl_2$ water bath. The vials contained varying macromolecular content, including 2%, 4%, and 8% agar solutions (weight per volume) prepared in DI water,[62] as well as boiled egg white.[63] The BTS scans used the following 3D imaging parameters: matrix size=128×128×12 with a 1.3 × 1.3 × 5 mm³ along the coronal orientation, TE/TR=12/80 ms, and excitation flip angles of 5°, 10°, 20°, 40°, and 60°, using a constant excitation pulse duration of 500 μs. An 8 ms Fermi pulse was applied with a peak $B_{1,\text{max}}$ of 9.128 $\mu T$ and a carrier frequency offset of 4000 Hz.

### 3.3 Implementations

REFINE-MORE was implemented using Python 3.10 and PyTorch 1.13.1 on a workstation with an Intel(R) Xeon(R) Gold 6338 CPU and an NVIDIA A100 GPU with 80 GB memory. The framework was trained using the Adam optimizer[64] and a fixed learning rate of 0.001. For the INR module, the input coordinates were normalized to the range of [0,1] to facilitate the INR learning. The hash encoding was realized based on tiny-cuda-nn (https://github.com/NVlabs/tiny-cuda-nn), a publicly available CUDA-based library. A lightweight MLP with 3 hidden layers and 64 neurons per layer was used in combination with ReLU activations for hidden layers and with no activation applied to the final output layer. In the physics reinforcement module, the proximal gradient descent algorithm was unrolled into a total of 4 phases. U-Net takes $N = 4$ quantitative maps ($I_0 = $ $\rho M_0, T_1^F, f, k_F$) as input channels, and the first layer outputs 32 feature maps. The model was trained progressively: the INR module was first trained to minimize $\mathcal{L}_1 + \mathcal{L}_2$ for 1500 iterations to generate a reasonable initialization, and then both the INR in the initialization module and the U-Net in the physics reinforcement module were jointly optimized to minimize $\mathcal{L}_1 + \mathcal{L}_2 + \mathcal{L}_3$ for an additional 1500 iterations. The regularization weight $\lambda$ in Eq. (4) was fixed at 0.01 across all experiments. The REFINE-MORE code is available at: https://github.com/I3Tlab/REFINE-MORE.

### 3.4 Experimental setup

We conducted the following experiments to evaluate the proposed REFINE-MORE method.

#### 3.4.1 Experiment 1: Method comparison

This experiment aimed to assess the performance of REFINE-MORE compared with other state-of-the-art methods. The in vivo BTS k-space data were retrospectively undersampled along two phase encoding directions using a 2D variable density Gaussian sampling pattern with acceleration factors (AF) of 4 and 5. The 24×24 center k-space region was fully sampled to calculate coil sensitivity maps using the ESPIRiT algorithm[65] (see **Supporting Information Figure S2** for detailed sampling masks). REFINE-MORE was compared with Zero Filling, locally low-rank (LLR) method,[13] MANTIS,[29] RELAX,[33] and SUMMIT.[34] For Zero Filling and LLR, the weighted images at different flip angles were first reconstructed, and then the multiparametric maps were estimated by the nonlinear fitting using the "fmincon" function in MATLAB (Mathworks, Natick, MA, USA). MANTIS is a supervised qMRI reconstruction method that performs end-to-end mapping from aliased weighted images to the corresponding quantitative maps, requiring fully sampled data as ground truth for training. In contrast, RELAX is a self-supervised approach that leverages the underlying MR signal model to directly estimate quantitative maps from undersampled data without the need for fully sampled references. To evaluate both methods, we conducted a leave-one-out five-fold cross-validation on the in vivo dataset, where in each fold, four subjects were used for training and the remaining one for testing. We further evaluated SUMMIT, an INR-based method that directly reconstructs quantitative maps without explicitly incorporating MR signal modeling. All methods were performed using the source code with default parameter settings from their original developers after adapting the MR signal model to BTS. The fully sampled data processed by nonlinear voxel-wise fitting served as the reference. The results of different methods were compared qualitatively and quantitatively.

The quantitative evaluation metrics are normalized root mean square error (nRMSE) and structural similarity index measure (SSIM), where the nRMSE is defined as follows:

$$\text{nRMSE} = \frac{\|\hat{q} - q\|_2}{\|q\|_2} \times 100\% \qquad (15)$$

where $\hat{q}$ is the reconstructed quantitative map and $q$ represents the corresponding reference. Additionally, the region-of-interest (ROI) analysis was conducted to assess the performance across selected anatomical regions, including the frontal white matter, splenium of corpus callosum, genu of corpus callosum, thalamus, and caudate nucleus. A Bland-Altman plot[66] was generated to evaluate the agreement between the reconstructed quantitative values and the fully sampled reference, providing insights into systematic bias and variability across methods.

### 3.4.2 Experiment 2: Evaluation of robustness and accuracy

This experiment aimed to assess the robustness of REFINE-MORE on different imaging objects and its accuracy in estimating multiple quantitative parameters under simplified but controlled imaging conditions. The phantom BTS k-space data were retrospectively undersampled along one phase encoding direction using a 1D variable density Gaussian sampling pattern with AF=4 and 24 calibration lines (see **Supporting Information Figure S2** for detailed sampling masks). REFINE-MORE was applied to reconstruct the undersampled data, and the resulting quantitative maps of $T_1^{\text{F}}$, $f$, and $k_{\text{F}}$ were compared to those derived from fully sampled data. Mean values within each vial were extracted and analyzed using linear regression to evaluate estimation accuracy.

### 3.4.3 Experiment 3: Model adaptation for improving computational efficiency

This experiment aimed to demonstrate the potential benefits of incorporating model adaptation strategies into the REFINE-MORE framework to reduce reconstruction time. Specifically, the learned weights of the hash encodings, MLPs, and U-Nets from previously reconstructed one subject were reused as the starting point when initializing the model for a new subject. To evaluate the impact, loss curves with and without model adaptation were plotted and compared. Additionally, the reconstructed quantitative maps at different iteration stages were visualized to illustrate the accelerated convergence.

### 3.4.4 Experiment 4: Other ablation studies

A series of ablation experiments were conducted to validate the design choices and effectiveness of key components in REFINE-MORE. To demonstrate the rationale of the proposed separate spatial and temporal encoding strategy for the weighted image reconstruction, we visualized the learned features from the hash encodings and performed temporal interpolation to verify that the temporal features capture reasonable contrast variations. In addition, we evaluated the impact of the physics reinforcement module by comparing models without it and with varying numbers of unrolled phases to investigate its effectiveness in improving the accuracy of multiparametric estimation. Finally, we explored the influence of the hyperparameter $\lambda$ in Eq. (4) on the reconstructed weighted images, confirming that the chosen value achieves a good balance between noise suppression and structural preservation in our datasets.

## 4. RESULTS

### 4.1 Results of different methods

**Figure 2** presents the reconstructed $I_0$, $T_1^F$, $f$, and $k_F$ maps by different methods on the in vivo data at AF=4. The top two rows are the reconstructed $I_0$ maps. Visually, the direct Zero Filling approach results in blurred reconstructions with noticeable artifacts. LLR and SUMMIT can partially suppress these artifacts, as illustrated by the zoomed-in images. The results of RELAX and MANTIS are over-smoothed, making it difficult to discern cortical structures. In contrast, REFINE-MORE yields the most faithful reconstruction that closely resembles the fully sampled reference. Regarding the quantitative $T_1^F$ map, REFINE-MORE demonstrates superior structure preservation, particularly at gray-white matter boundaries, and maintains consistent contrast across the white matter. The other methods either oversmooth the tissue contrast (e.g., RELAX and MANTIS) or retain noise-like variations (e.g., Zero Filling, LLR, and SUMMIT). For $f$ and $k_F$ maps, REFINE-MORE exhibits superior visual agreement with the fully sampled results, capturing quantitative variations more accurately and yielding spatial patterns that closely resemble the reference.

**Figure 3** shows the reconstruction results of different methods at a 5× 2D variable density Gaussian undersampling pattern. Similar to the AF=4 case, the $I_0$ map reconstructed by REFINE-MORE demonstrates effective artifact suppression and preserves fine structural details, as highlighted in the zoomed-in regions. For the $T_1^F$ map, methods such as Zero Filling, RELAX, and MANTIS produce blurred reconstructions, which obscure anatomical boundaries and compromise quantitative accuracy. In comparison, LLR, SUMMIT, and REFINE-MORE produce maps more consistent with references, although LLR and SUMMIT exhibit higher levels of residual noise.

Regarding the reconstructed $f$ maps, LLR, MANTIS, and REFINE-MORE achieve better contrast alignment with the reference, whereas Zero Filling, RELAX, and SUMMIT tend to underestimate the values. For the $k_F$ map, REFINE-MORE provides the closest match to the reference in terms of contrast fidelity.

**Table 1** summarizes nRMSE and SSIM values comparing the reconstructions from different methods with the fully sampled reference. REFINE-MORE consistently achieves the lowest nRMSE and the highest SSIM across various acceleration factors and quantitative parameter maps, except for the $k_F$ map at AF = 4, where its SSIM is 0.3 lower than the best-performing method. In contrast, RELAX and MANTIS exhibit inferior quantitative performance, likely due to limited training data, which may hinder their ability to learn representative features for accurate reconstruction. **Figure 4** displays the Bland-Altman plots for Zero Filling, LLR, SUMMIT, and REFINE-MORE across five ROIs. In Bland-Altman plots, the horizontal axis represents the mean of the reconstructed values and the reference values, and the vertical axis shows the errors of the reconstructed results and the reference. Therefore, we can see that LLR and REFINE-MORE exhibit smaller mean biases and narrower limits of agreement in the reconstruction of $T_1^F$ and $f$, while for the $k_F$ map, REFINE-MORE achieves the lowest mean bias and the tightest limits of agreement, indicating good overall consistency with the reference across all three maps. The plots for RELAX and MANTIS are provided separately in **Supporting Information Figure S3** due to larger-scale errors likely caused by small training data.

### 4.2 Results on phantom data

The results of phantom data are depicted in **Figure 5**. **Figure 5(A)** shows the $I_0$, $T_1^F$, $f$, and $k_F$ maps reconstructed by Zero Filling, REFINE-MORE, and Fully Sample. The quantitative maps of Zero Filling exhibit ripple artifacts due to the undersampling, whereas these artifacts can be effectively eliminated by the REFINE-MORE method. **Figure 5(B)** illustrates the linear regression analysis between the reconstructed values from Zero Filling and REFINE-MORE and the fully sampled reference. The mean $T_1^F$, $f$, and $k_F$ values of the four vials estimated by REFINE-MORE demonstrate good agreement with those estimated from the fully sample data, as indicated by the correlation slope approaching 1 with $R^2$=1.000, 0.999, and 0.999, respectively. In contrast, the Zero Filling reconstructions show pronounced deviations from the reference values, mainly attributed to artifacts that impair quantitative accuracy.

## 4.3 Results of model adaptation

**Figure 6** illustrates the advantages of the model adaptation strategy. **Figure 6(A)** demonstrates the benefits of reusing the temporal features and MLP weights learned from one subject for a new subject during the initialization module. As shown by the loss convergence curves, the loss decreases faster and reaches a lower value when this model adaptation strategy is employed. The right panel of **Figure 6(A)** presents the reconstruction results at 200, 500, and 1500 iterations, with and without model adaptation, respectively. Notably, the adapted model achieves comparable reconstruction accuracy at 500 iterations to that of the non-adapted model at 1500 iterations. **Figure 6(B)** demonstrates the benefits of LoRA-based model adaptation during the physics reinforcement module. It can be observed that the adapted model exhibits faster convergence, whereas the non-adapted model shows greater fluctuations in loss. This is further supported by comparing the quantitative maps generated with and without model adaptation at 100, 200, and 1500 iterations. With adaptation, the model reaches a similar level of performance in 200 iterations, which results in an approximately five-fold improvement in reconstruction efficiency (with adaptation: (6.6+7.6) min vs. without adaptation: (13.3+58.5) min). These results highlight the effectiveness of the proposed model adaptation strategies in improving reconstruction accuracy and efficiency.

## 4.4 Results of ablation studies

**Figure 7** visualizes spatial and temporal features learned by the hash encodings $\mathcal{H}_1$ and $\mathcal{H}_2$, respectively. **Figure 7(A)** shows representative spatial features at five different spatial resolution levels. It is observed that the spatial features evolve from low-level, global contrast representations to higher-level, more abstract structural features as the resolution increases. This progression highlights the hierarchical encoding capability of the hash-based spatial representation, enabling the model to capture both coarse and fine-grained image characteristics. **Figure 7(B)** presents the temporal features across 8 time points corresponding to 4 different flip angles with and without off-resonance saturation, respectively, in the BTS protocol. To further validate that the temporal features indeed encode contrast variations, we generated three interpolated temporal features at $t = 2.5, 3.5,$ and $4.5$, which were not involved during the model optimization. When concatenated with the learned spatial features and used to query the MLP, they produce images exhibiting distinct contrast weightings. These results demonstrate the model's ability to generalize to intermediate temporal states and confirm the semantic continuity encoded in the temporal feature space.

**Figure 8** illustrates the reconstruction quality of $T_1^F$ and $f$ maps under different configurations of the unrolled physics reinforcement module, including the case without reinforcement and with varying numbers of total unrolling phases. In the absence of reinforcement, the $T_1^F$ map exhibits overly smoothed contrast, and the $f$ map shows noticeable inaccuracies compared to the fully sampled reference in the corpus callosum region. As the number of unrolling phases increases, the $T_1^F$ maps progressively recover finer tissue details, as shown in the zoomed-in images. When the number of unrolling phases exceeds 3, further improvements become marginal, suggesting that the performance gains reach a saturation point. Based on this observation, using 4 unrolling phases offers a favorable trade-off between reconstruction quality and computational cost in our datasets.

**Figure 9** presents the weighted images reconstructed during the initialization module using different $\lambda$ values in the $\mathcal{L}_1$ loss term. The zoomed-in images reveal that as $\lambda$ increases, the reconstructed images become progressively smoother, with reduced noise levels. However, excessively large $\lambda$ values lead to over-smoothing, resulting in the loss of fine structural details. Based on this observation, we selected $\lambda=0.01$ as it provides a good trade-off between noise suppression and structural preservation in our datasets.

## 5. DISCUSSION AND CONCLUSION

In this study, we proposed a self-supervised and scan-specific framework, REFINE-MORE, for accelerated multiparametric qMRI reconstruction. The method first employs the concept of implicit neural representation to model the entire 4D volumes as functions of coordinates, providing an initial estimate of multiple quantitative maps. This initialization is then refined using an unrolled proximal gradient descent CNN framework, which enforces data consistency with the measured k-space data through physics-based modeling. We validated the proposed method on the datasets acquired by a multiparametric magnetization transfer imaging sequence for simultaneous reconstruction of $T_1^F$, $f$, and $k_F$ maps. Experimental results demonstrate that REFINE-MORE achieves accurate and robust multiparametric estimation with 25% and 20% k-space data. Furthermore, incorporating model adaptation strategies significantly improves reconstruction efficiency, reducing the overall reconstruction time by fivefold. In the following subsections, we further discuss REFINE-MORE from several key perspectives.

## 5.1 Analysis of implicit neural representation-based initialization

In this study, we adopted INR to address the challenges of multiparametric qMRI reconstruction, offering a fundamentally different strategy from conventional CNN-based methods. While CNNs typically operate on artifact-laden images using convolutional filters to suppress artifacts and recover quantitative maps, INR directly models the target image as a continuous function of coordinates, providing several key advantages. First, leveraging recent advances in INR, we combine multi-resolution hash encoding with an MLP as the representation function. Efficient CUDA-based implementations of hash encodings make this approach both memory-efficient and computationally fast, enabling practical application to high-dimensional qMRI data. As a result, our method can handle entire four-dimensional datasets without resorting to slice-wise processing. For example, our INR module jointly reconstructs a 176×176×48×8 weighted image series together with four 176×176×48 quantitative parameter maps ($I_0$, $T_1^F$, $f$, and $k_F$), using only 49 GB of memory. Second, MLPs, as universal function approximators,[67] can reliably fit complex continuous functions under a differentiable loss, providing a stable and flexible reconstruction framework. Third, the coordinate-based formulation allows incorporation of prior knowledge. By separately encoding spatial and temporal coordinates, the INR implicitly models spatiotemporal correlations, helping the network capture the underlying data structure, as demonstrated in **Figure 7**. These advantages make INR a powerful tool, producing high-quality parameter estimates that serve as effective inputs for the subsequent physics-based reinforcement stage.

## 5.2 Effectiveness of physics reinforcement

The physics reinforcement module plays a crucial role in refining the initial estimates produced by the INR, resulting in more realistic parameter maps. In our framework, this module is implemented as a model-based optimization process, specifically using an unrolled proximal gradient descent algorithm. Comparisons with the SUMMIT method (**Figures 2** and **3**) and the ablation study (**Figure 8**) highlight the effectiveness of this module. This improvement is likely because INR, relying solely on loss-based supervision, may produce biased or overly smoothed results due to regularization, such as the piecewise constant patterns observed in the $f$ maps reconstructed by SUMMIT. In contrast, the physics reinforcement module explicitly corrects such deviations by aligning the reconstruction with the measured k-space data through known forward signal and acquisition models. This ensures that even if the INR initial estimates are imperfect, the final output remains consistent with the true underlying tissue properties.

### 5.3 Benefits of model adaptation for scan-specific methods

REFINE-MORE is a scan-specific reconstruction framework that requires no external training datasets beyond the target scan itself, making it particularly advantageous when fully sampled data are impractical to acquire. As shown in our experiments, methods like RELAX and MANTIS exhibit degraded performance under such data-limited conditions. However, scan-specific methods have inherent limitations, as they optimize a separate model for each scan and cannot leverage shared priors across subjects or datasets. Unlike supervised learning, which benefits from diverse training examples, scan-specific frameworks often ignore reusable knowledge, potentially wasting informative representations. To address this, we introduce a model adaptation strategy to transfer learned priors from previously reconstructed scans to new target data. For example, we apply LoRA-based adaptation, training only a small number of additional parameters while freezing the core network weights. This reduces trainable parameters from 7,763,879 in the original U-Net to 517,600. As demonstrated in **Figure 6(B)**, LoRA adaptation achieves comparable performance in roughly one-fifth of the original training time. These results suggest that incorporating prior knowledge can enhance performance and training efficiency in scan-specific frameworks, and this strategy may be extended beyond REFINE-MORE.

### 5.4 Limitations

First, REFINE-MORE does not directly exploit population-level priors or anatomical consistency across subjects, which supervised frameworks may leverage to further stabilize reconstruction. Second, like conventional iterative optimization, this method may need careful adaptation or hyperparameter tuning to extend to other contrasts or anatomies with different signal models. Third, while techniques such as LoRA have been incorporated to accelerate REFINE-MORE, the per-scan optimization remains relatively time-consuming compared to pretrained supervised methods that perform inference within seconds.

### ACKNOWLEDGEMENTS

The research reported in this publication was supported by the National Institute of Biomedical Imaging and Bioengineering under Award Number R21EB031185, the National Institute of Arthritis and Musculoskeletal and Skin Diseases under Award Numbers R01AR081344, R01AR079442, and R56AR081017.

# LIST OF TABLES

**Table 1.** Comparison of nRMSE and SSIM values relative to the fully sampled reference for different methods on the in vivo human brain dataset at acceleration factors (AF) of 4 and 5. The results are reported as mean ± standard deviation. The best-performing method in each category is highlighted in bold.

|  | Mean±SD at AF=4 | | | | | |
|--|--|--|--|--|--|--|
|  | $T_1^F$ | | $f$ | | $k_F$ | |
|  | nRMSE (%) | SSIM (%) | nRMSE (%) | SSIM (%) | nRMSE (%) | SSIM (%) |
| Zero Filling | 21.5±1.8 | 95.6±0.5 | 25.0±1.5 | 94.8±0.8 | 25.1±1.5 | 94.9±0.8 |
| LLR | 19.6±2.8 | 96.5±0.8 | 20.9±2.4 | 95.6±1.2 | 24.2±3.4 | 94.9±0.8 |
| RELAX | 31.2±4.9 | 94.3±0.8 | 27.4±1.5 | 93.5±0.6 | 27.8±2.1 | 93.4±0.7 |
| MANTIS | 27.5±5.4 | 94.7±1.1 | 26.3±2.5 | 95.1±0.8 | 28.2±3.1 | 94.7±1.0 |
| SUMMIT | 19.2±4.2 | 96.7±1.1 | 20.7±3.9 | 96.4±1.0 | 20.8±3.9 | **96.4±0.9** |
| REFINE-MORE | **15.4±2.1** | **98.1±0.4** | **20.1±3.4** | **96.6±0.9** | **20.5±3.5** | 96.1±0.3 |
|  | Mean±SD at AF=5 | | | | | |
|  | $T_1^F$ | | $f$ | | $k_F$ | |
|  | nRMSE (%) | SSIM (%) | nRMSE (%) | SSIM (%) | nRMSE (%) | SSIM (%) |
| Zero Filling | 23.2±1.9 | 95.1±0.5 | 26.5±1.4 | 94.2±0.7 | 26.6±1.4 | 94.3±0.7 |
| LLR | 21.4±2.8 | 95.9±0.8 | 22.6±2.3 | 94.9±1.2 | 25.9±3.2 | 94.2±0.9 |
| RELAX | 31.1±4.0 | 94.1±0.8 | 30.6±0.6 | 92.5±0.6 | 31.0±1.3 | 92.4±0.8 |
| MANTIS | 28.3±4.5 | 94.7±1.1 | 28.4±4.0 | 94.6±1.0 | 30.0±4.3 | 94.3±1.3 |
| SUMMIT | 21.3±4.6 | 95.9±1.3 | 22.8±4.1 | 95.7±1.0 | 22.9±4.1 | 95.7±1.1 |
| REFINE-MORE | **17.2±2.7** | **97.6±0.6** | **22.0±4.1** | **96.1±1.0** | **22.1±3.6** | **96.1±1.0** |

# FIGURE CAPTIONS

**Figure 1.** Overview of the proposed REFINE-MORE framework, which adopts a two-stage optimization strategy. (A) The initialization module represents the weighted images as a function of spatial and temporal coordinates, parameterized by hash encodings $\mathcal{H}_1$ and $\mathcal{H}_2$, and the $MLP_1$, while the multiparametric maps are represented as a function of spatial coordinates, parameterized by the hash encoding $\mathcal{H}_3$ and the $MLP_2$. The parameters in hash encodings and MLPs are optimized by minimizing the loss functions $\mathcal{L}_1$ and $\mathcal{L}_2$. (B) The physics reinforcement module unrolls the physics model-based proximal gradient descent algorithm into $K$ phases and uses a U-Net as an implicit proximal operator. The U-Net is optimized by minimizing the data consistency loss $\mathcal{L}_3$ applied to the outputs of all phases. This module starts with the initial estimate and progressively refines the quantitative maps. The final quantitative maps are obtained from the last phase.

**Figure 2.** Qualitative comparison of $I_0$, $T_1^F$, $f$, and $k_F$ maps reconstructed by different methods at AF=4. The regions enclosed by white boxes are shown in enlarged views. Visually, REFINE-MORE yields the most faithful reconstruction, with sharper anatomical boundaries and removed artifacts that closely resemble the fully sampled reference.

**Figure 3.** Qualitative comparison of $I_0$, $T_1^F$, $f$, and $k_F$ maps reconstructed by different methods at AF=5. The regions enclosed by white boxes are shown in enlarged views. Similar to the AF=4 case, REFINE-MORE exhibits superior visual agreement with the fully sampled results across different maps.

**Figure 4.** Bland-Altman plots for (A) Zero Filling (ZF), (B) LLR, (C) SUMMIT, and (D) REFINE-MORE across five ROIs. The solid lines indicate mean differences, and the dashed lines represent the 95% confidence level.

**Figure 5.** Presentation of the phantom experiment results at AF=4. (A) $I_0$, $T_1^F$, $f$, and $k_F$ maps reconstructed by Zero Filling, REFINE-MORE, and Fully Sample. The quantitative maps of Zero Filling exhibit ripple artifacts, whereas these artifacts can be effectively eliminated by REFINE-MORE. (B) Linear regression analysis between the reconstructed values from Zero Filling and REFINE-MORE and the fully sampled reference. The red line represents the linear fitting, and the black dashed line corresponds to $y = x$. The regression slope between REFINE-MORE and the reference is closer to 1.

**Figure 6.** Illustration of the benefits of the model adaptation strategy. (A) The left panel shows the loss curves of the models with and without adaptation during the initialization module. The right panel displays the reconstructed quantitative maps at 200, 500, and 1500 iterations. (B) The left panel shows the loss curves of the models with and without adaptation during the physics reinforcement module. The right panel presents the reconstructed quantitative maps at 100, 200, and 1500 iterations. The corresponding nRMSE relative to the fully sampled reference is indicated in each image.

**Figure 7.** Visualization of spatial and temporal features learned by the hash encoding $\mathcal{H}_1$ and $\mathcal{H}_2$. (A) Representative spatial features at five different resolution levels. (B) Display of the temporal features across 8 time points. (C) The synthesized images with different contrasts at time points of 2.5, 3.5, and 4.5.

**Figure 8.** Reconstructed $T_1^F$ and $f$ maps without the physics reinforcement module and with varying numbers of total unrolling phases. The regions enclosed by white boxes are shown in enlarged views.

**Figure 9.** Effects of $\lambda$ on the reconstructed weighted images.

**Supporting Background Information.** An Overview of Hash Encoding.

**Supporting Information Figure S1.** Illustration of the model adaptation strategy. (A) For the initialization module, we reuse and freeze the temporal features and MLP weights from previously reconstructed data. (B) For the physics reinforcement module, we incorporate a low-rank adaptation (LoRA) strategy that freezes most parameters in U-Net and only learns a small set of parameters.

**Supporting Information Figure S2.** Display of undersampling masks for in vivo and phantom experiments. (A) 2D variable density Gaussian undersampling patterns with AF=4 and AF=5 for in vivo experiments. (B) A 1D variable density Gaussian undersampling pattern with AF=4 for phantom experiments.

**Supporting Information Figure S3.** Bland-Altman plots for (A) RELAX and (B) MANTIS across five ROIs. The solid lines indicate mean differences, and the dashed lines represent the 95% confidence level.

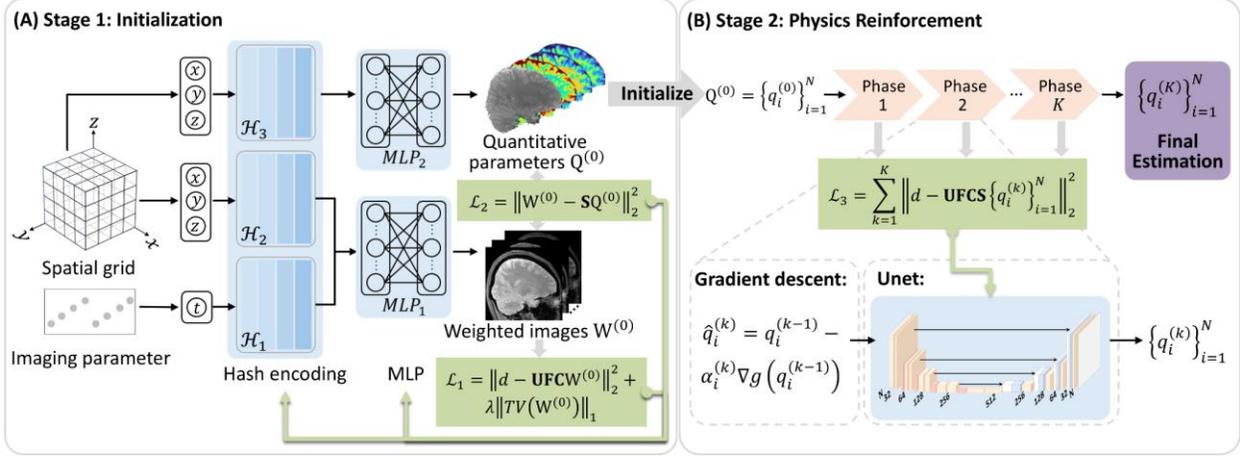

**Figure 1.** Overview of the proposed REFINE-MORE framework, which adopts a two-stage optimization strategy. (A) The initialization module represents the weighted images as a function of spatial and temporal coordinates, parameterized by hash encodings $\mathcal{H}_1$ and $\mathcal{H}_2$, and the $MLP_1$, while the multiparametric maps are represented as a function of spatial coordinates, parameterized by the hash encoding $\mathcal{H}_3$ and the $MLP_2$. The parameters in hash encodings and MLPs are optimized by minimizing the loss functions $\mathcal{L}_1$ and $\mathcal{L}_2$. (B) The physics reinforcement module unrolls the physics model-based proximal gradient descent algorithm into $K$ phases and uses a U-Net as an implicit proximal operator. The U-Net is optimized by minimizing the data consistency loss $\mathcal{L}_3$ applied to the outputs of all phases. This module starts with the initial estimate and progressively refines the quantitative maps. The final quantitative maps are obtained from the last phase.

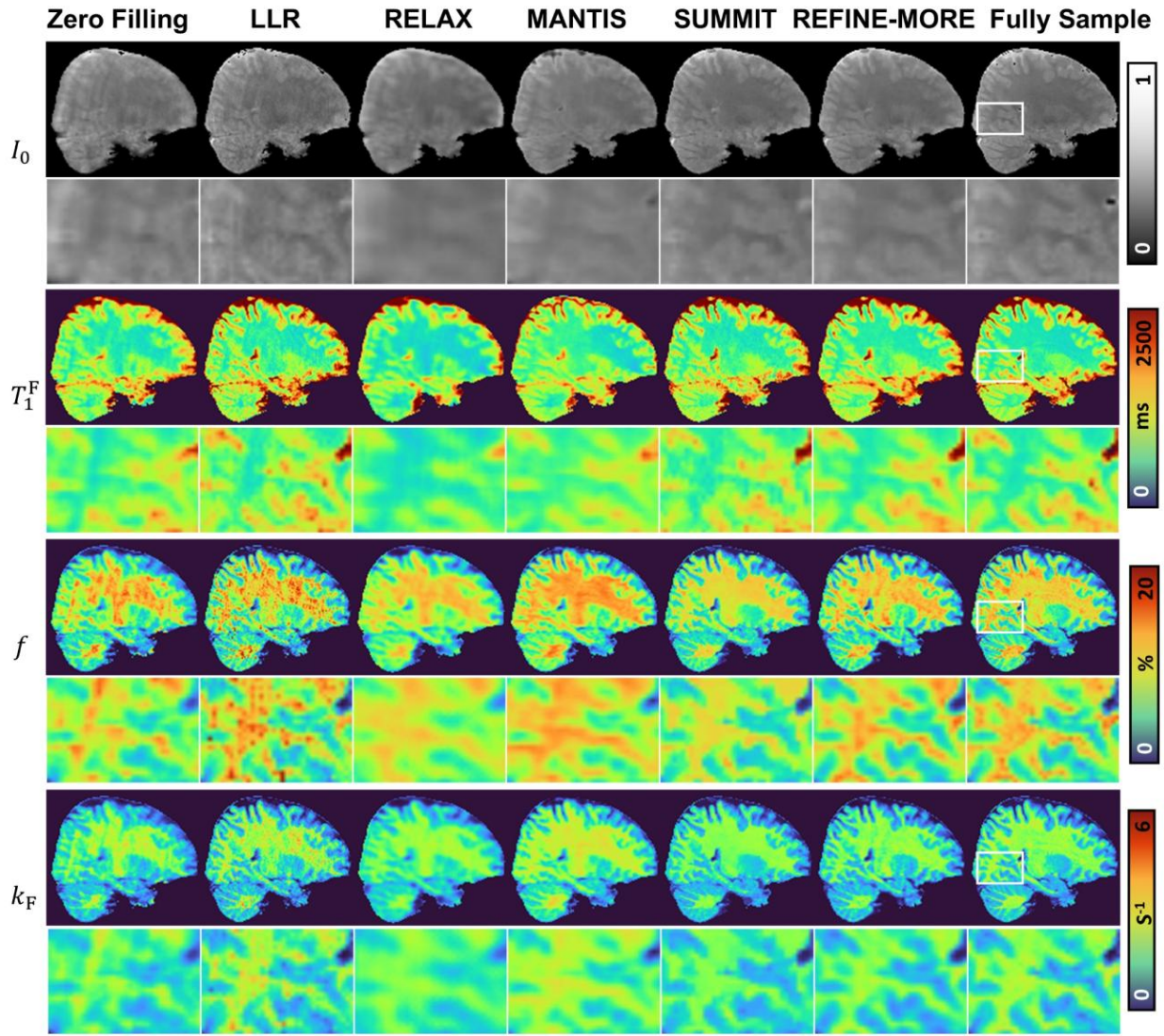

**Figure 2.** Qualitative comparison of $I_0$, $T_1^F$, $f$, and $k_F$ maps reconstructed by different methods at AF=4. The regions enclosed by white boxes are shown in enlarged views. Visually, REFINE-MORE yields the most faithful reconstruction, with sharper anatomical boundaries and removed artifacts that closely resemble the fully sampled reference.

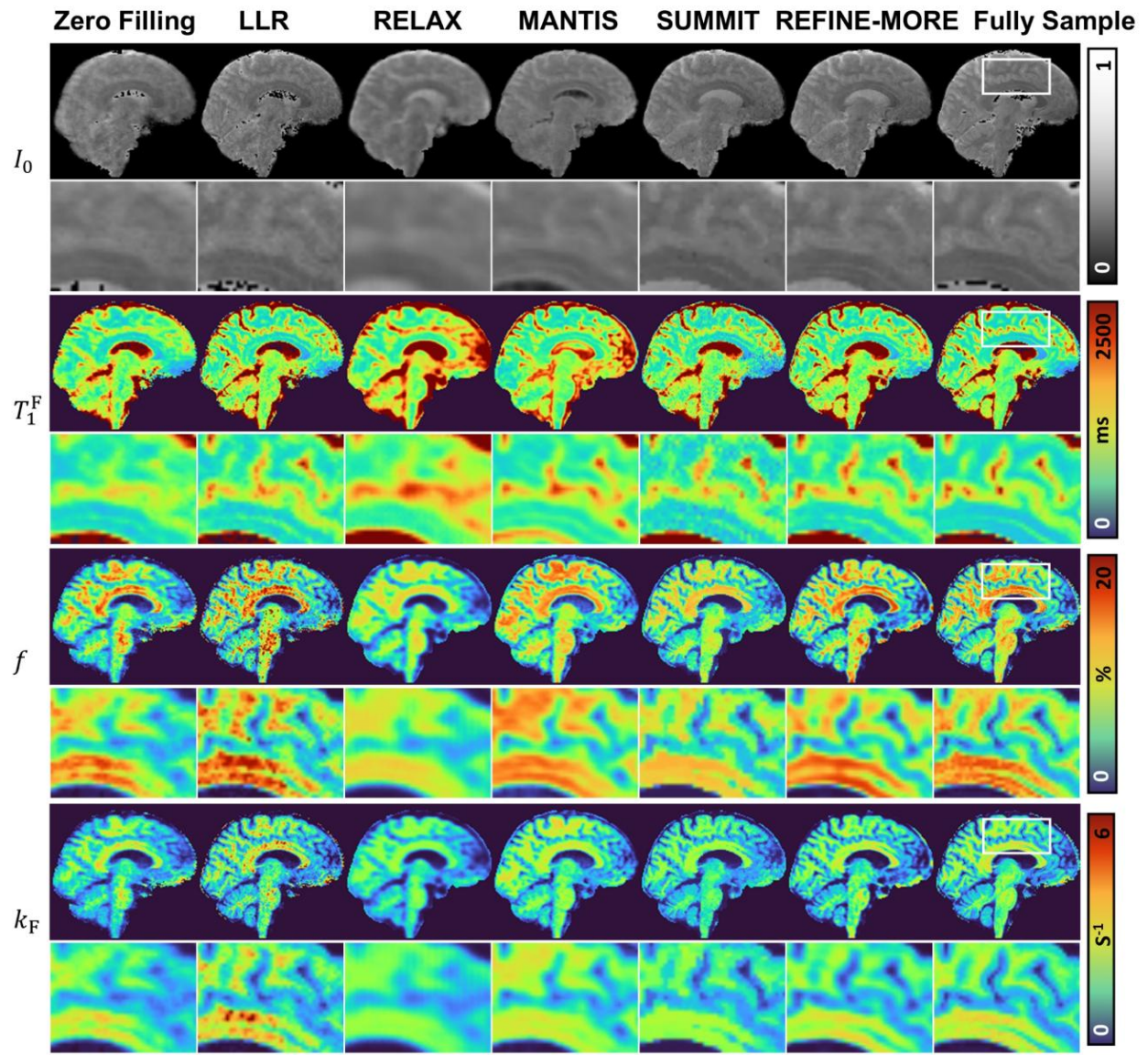

**Figure 3.** Qualitative comparison of $I_0$, $T_1^F$, $f$, and $k_F$ maps reconstructed by different methods at AF=5. The regions enclosed by white boxes are shown in enlarged views. Similar to the AF=4 case, REFINE-MORE exhibits superior visual agreement with the fully sampled results across different maps.

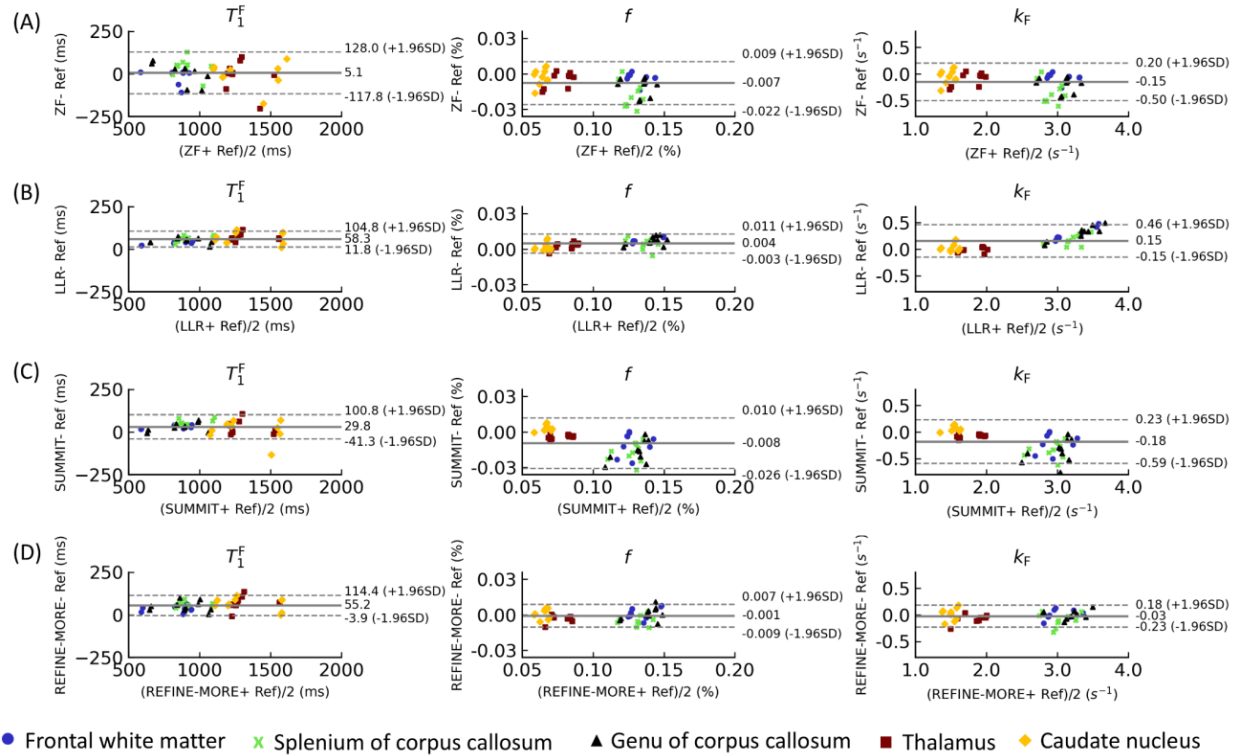

**Figure 4.** Bland-Altman plots for (A) Zero Filling (ZF), (B) LLR, (C) SUMMIT, and (D) REFINE-MORE across five ROIs. The solid lines indicate mean differences, and the dashed lines represent the 95% confidence level.

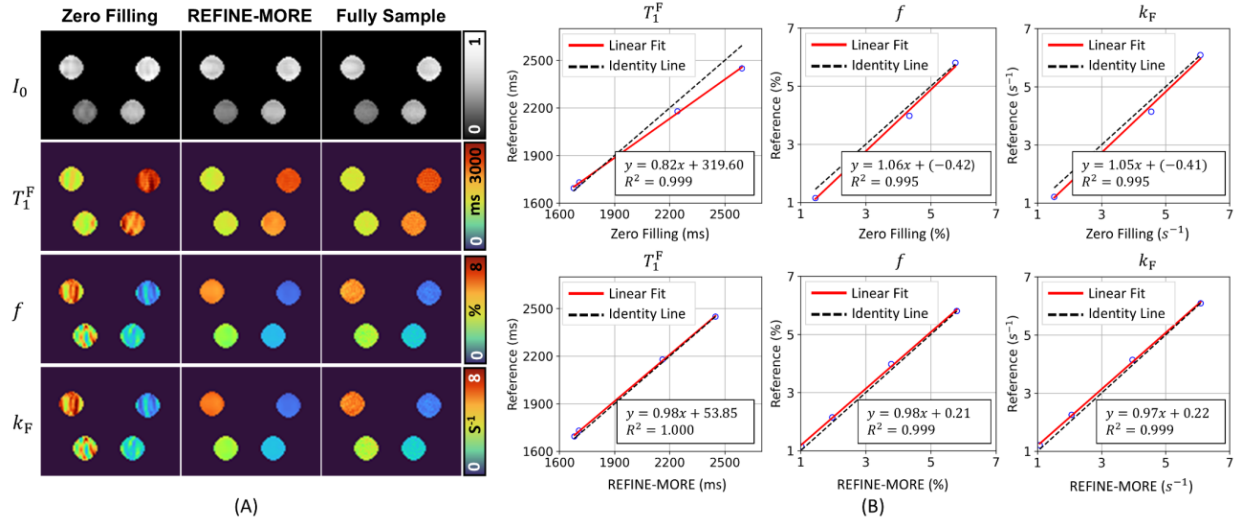

**Figure 5.** Presentation of the phantom experiment results at AF=4. (A) $I_0$, $T_1^F$, $f$, and $k_F$ maps reconstructed by Zero Filling, REFINE-MORE, and Fully Sample. The quantitative maps of Zero Filling exhibit ripple artifacts, whereas these artifacts can be effectively eliminated by REFINE-MORE. (B) Linear regression analysis between the reconstructed values from Zero Filling and REFINE-MORE and the fully sampled reference. The red line represents the linear fitting, and the black dashed line corresponds to $y = x$. The regression slope between REFINE-MORE and the reference is closer to 1.

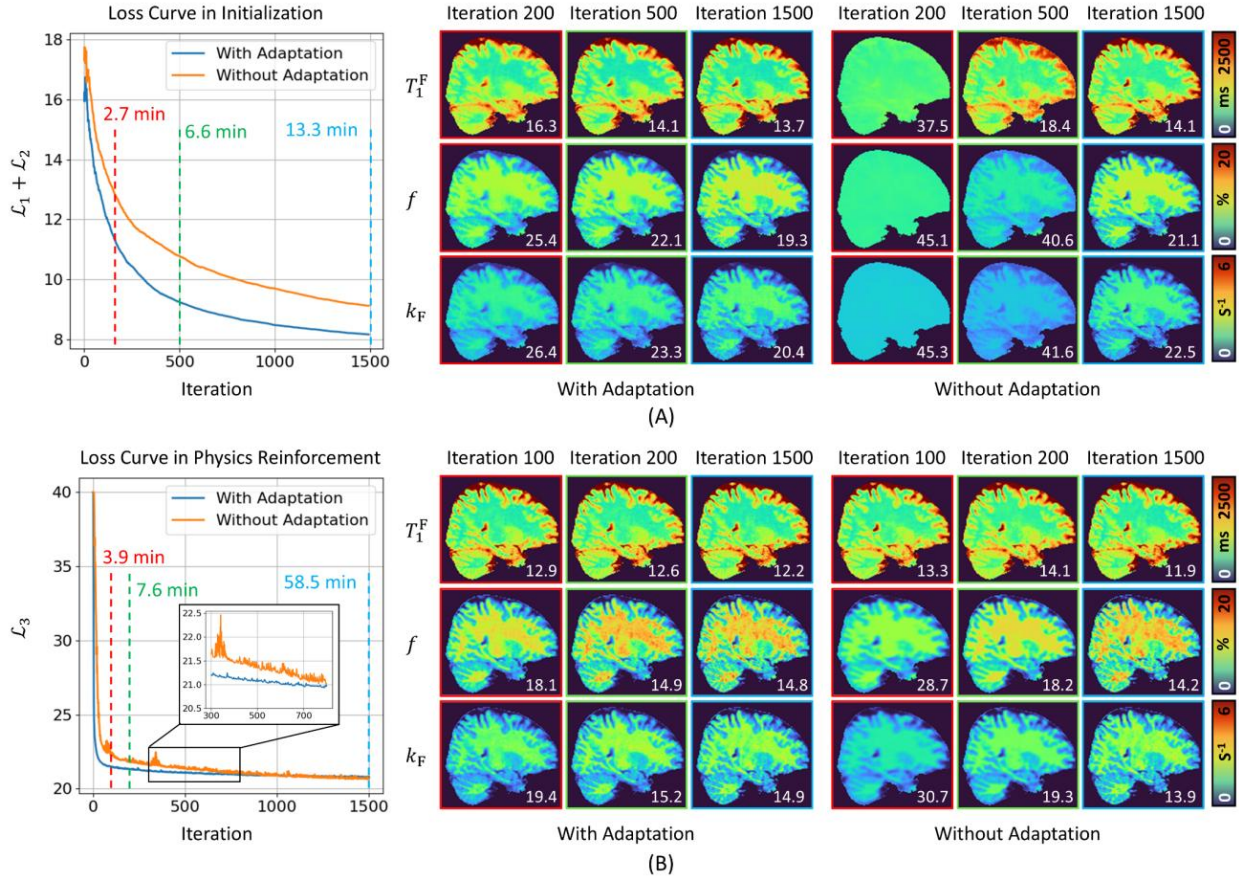

**Figure 6.** Illustration of the benefits of the model adaptation strategy. (A) The left panel shows the loss curves of the models with and without adaptation during the initialization module. The right panel displays the reconstructed quantitative maps at 200, 500, and 1500 iterations. (B) The left panel shows the loss curves of the models with and without adaptation during the physics reinforcement module. The right panel presents the reconstructed quantitative maps at 100, 200, and 1500 iterations. The corresponding nRMSE relative to the fully sampled reference is indicated in each image.

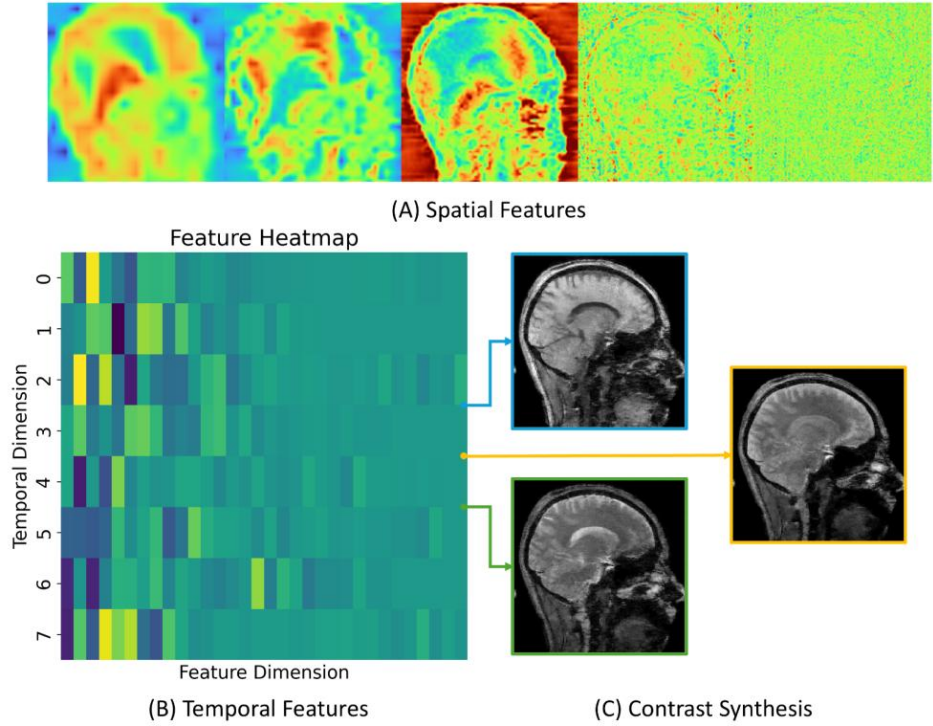

**Figure 7.** Visualization of spatial and temporal features learned by the hash encoding $\mathcal{H}_1$ and $\mathcal{H}_2$. (A) Representative spatial features at five different resolution levels. (B) Display of the temporal features across 8 time points. (C) The synthesized images with different contrasts at time points of 2.5, 3.5, and 4.5.

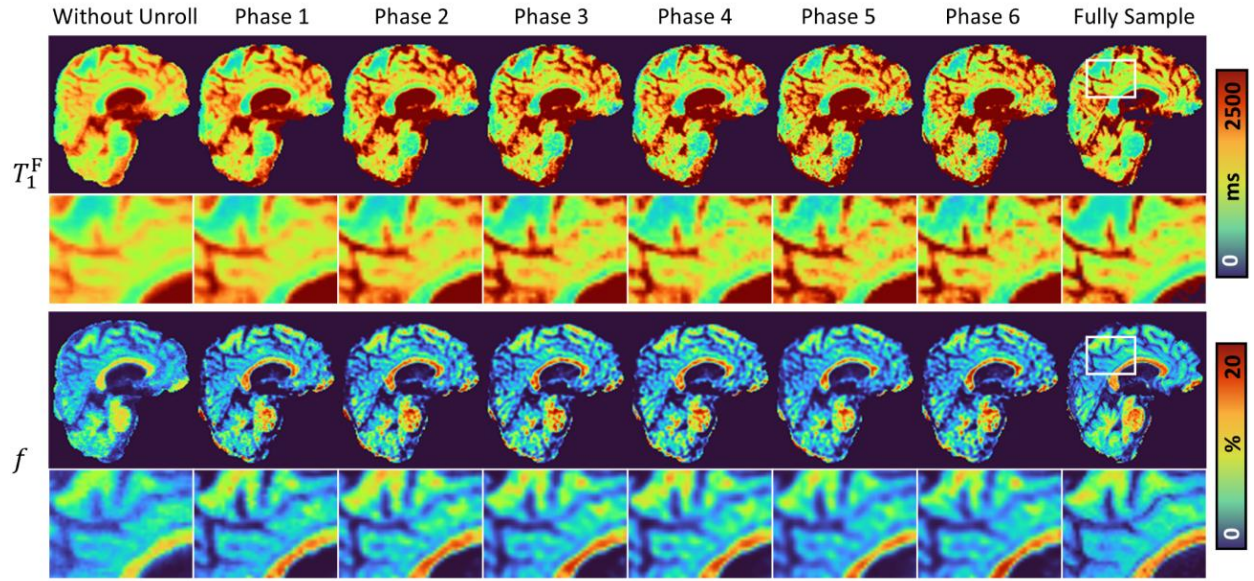

**Figure 8.** Reconstructed $T_1^F$ and $f$ maps without the physics reinforcement module and with varying numbers of total unrolling phases. The regions enclosed by white boxes are shown in enlarged views.

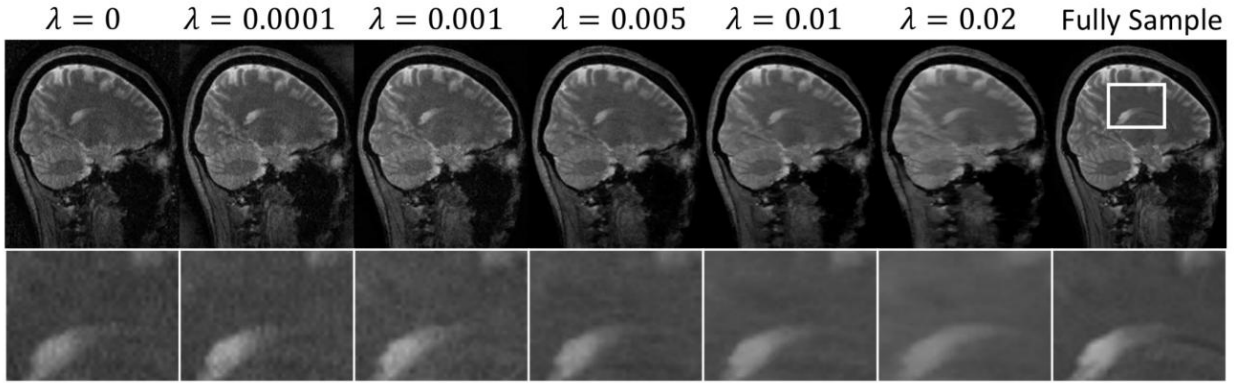

**Figure 9.** Effects of $\lambda$ on the reconstructed weighted images.

# Supporting Information

**Supporting Background Information.** An Overview of Hash Encoding.

**Hash encoding with multilayer perceptron as the representation function**

In this study, we employ a hash encoding[1] and an MLP as the continuous function to represent the images to be solved. Hash encoding maps the low-dimensional coordinates into a high-dimensional feature space, thereby facilitating the learning of high-frequency details. Compared to other encoding strategies in INR, hash encoding incorporates learnable parameters that permit the use of a much smaller MLP architecture while maintaining high representational capacity. This leads to faster convergence during training and improves memory efficiency, making it feasible to model and process entire 4D volumes on standard GPU hardware. Specifically, the learnable parameters in hash encoding are structured across $L$ distinct resolution levels, each implemented as a separate hash table. These levels span a range of geometrically increasing resolutions, starting from a base resolution $N_{min}$ and scaling by a factor of $b$ at each successive level. As a result, the resolutions follow the sequence $N_{min}, b \times N_{min}, \ldots, b^{L-1} \times N_{min}$. Each resolution level is composed of $T$ feature entries, and the dimensionality of each feature vector is $F$. Therefore, hash encoding provides a multi-resolution representation that enables hierarchical feature learning. Benefitting from the fast convergence of hash encoding combined with MLPs, INR can rapidly provide reasonable initialization estimates of 3D quantitative maps. The hyperparameters in hash encoding were fixed as the default values. For a more comprehensive description of hash encoding, readers are referred to the original paper.

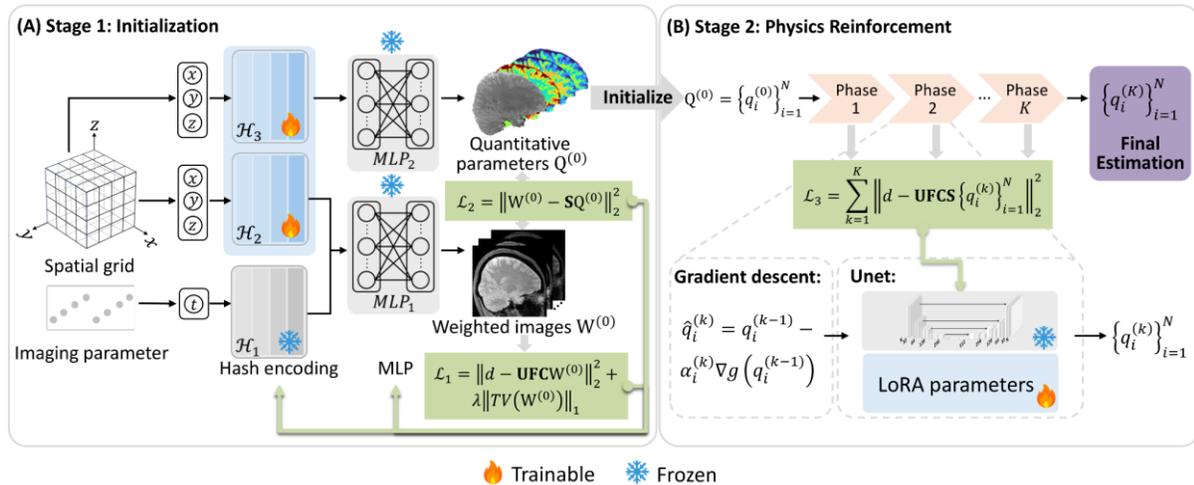

**Supporting Information Figure S1.** Illustration of the model adaptation strategy. (A) For the initialization module, we reuse and freeze the temporal features and MLP weights from previously reconstructed data. (B) For the physics reinforcement module, we incorporate a low-rank adaptation (LoRA) strategy that freezes most parameters in U-Net and only learns a small set of parameters.

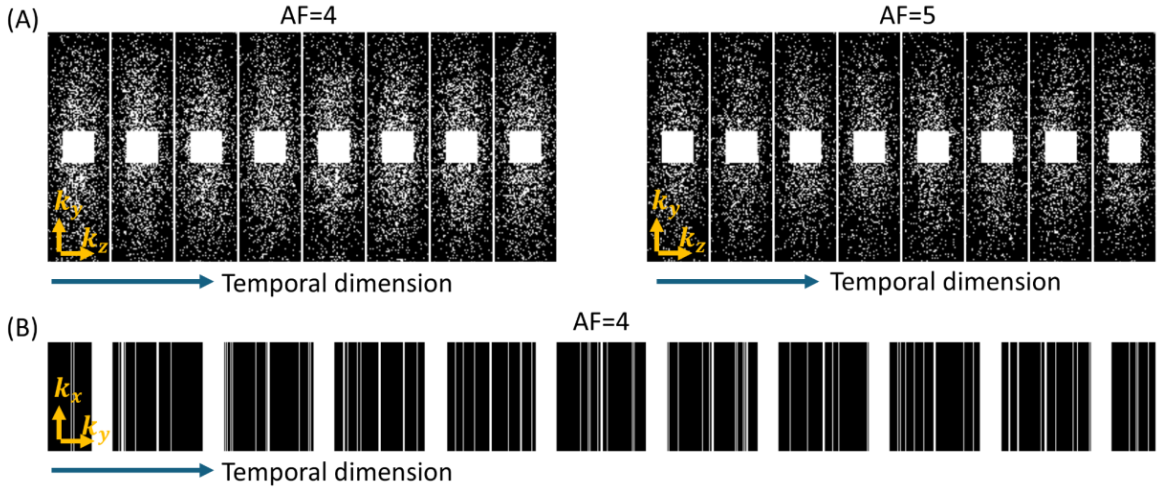

**Supporting Information Figure S2.** Display of undersampling masks for in vivo and phantom experiments. (A) 2D variable density Gaussian undersampling patterns with AF=4 and AF=5 for in vivo experiments. (B) A 1D variable density Gaussian undersampling pattern with AF=4 for phantom experiments.

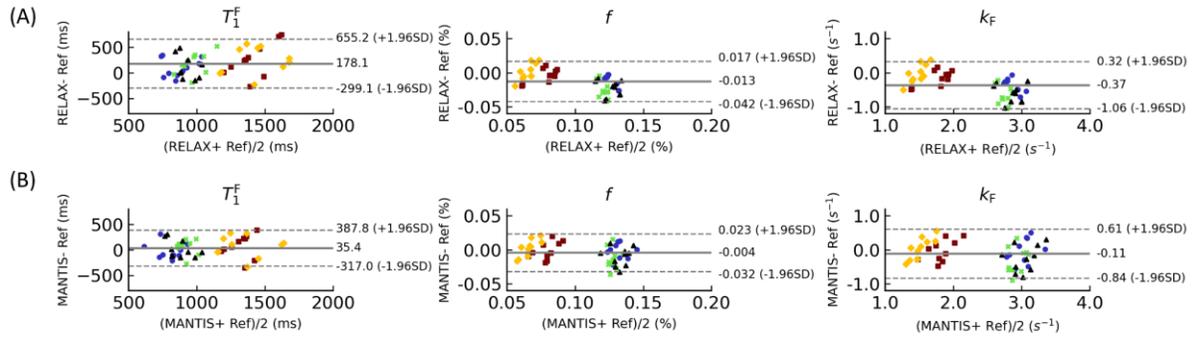

**Supporting Information Figure S3.** Bland-Altman plots for (A) RELAX and (B) MANTIS across five ROIs. The solid lines indicate mean differences, and the dashed lines represent the 95% confidence level.